\documentclass[aps,pra,amsmath,amssymb,reprint,groupedaddress,floatfix]{revtex4-2}
\newcommand{\bol}[1]{\boldsymbol{#1}}
\usepackage{physics}
\usepackage{tikz-cd}
\usepackage{tikz}
\usepackage{balance}
\usepackage{calc}
\usepackage{mathtools}
\usepackage{natbib}
\usepackage[pdftex,pdfpagelabels,bookmarks,hyperindex,hyperfigures,colorlinks]{hyperref}
\begin{document}

%\preprint{}

\title{Dyson Maps and Unitary Evolution for Maxwell Equations in Tensor Dielectric Media}


\author{Efstratios Koukoutsis}
\email{stkoukoutsis@mail.ntua.gr}
\author{Kyriakos Hizanidis}
%homepage[jfbjdjdb]{Your web page}
%\thanks{}
\affiliation{School of Electrical and Computer Engineering, National Technical University of Athens, Zographou 15780, Greece}

\author{Abhay K. Ram}
\affiliation{Plasma Science and Fusion Center, Massachusetts Institute of Technology, Cambridge,
Massachusetts 02139, USA}%

\author{George Vahala}
\affiliation{Department of Physics, William \& Mary, Williamsburg, VA 23187, USA}%


\date{\today}

\begin{abstract}
The propagation and scattering of electromagnetic waves in dielectric media is of theoretical and experimental interest in a wide variety of fields. An understanding of observational results generally requires a numerical solution of Maxwell equations -- usually implemented on conventional computers using sophisticated numerical algorithms. In recent years, advances in quantum information science and in the development of quantum computers have piqued curiosity about taking advantage of these resources for an alternate numerical approach to Maxwell equations. This requires a reformulation of the classical Maxwell equations into a form suitable for quantum computers which, unlike conventional computers, are limited to unitary operations.  In this paper, a unitary framework is developed for the propagation of electromagnetic waves in a spatially inhomogeneous, passive, non-dispersive, and anisotropic dielectric medium. For such a medium, generally, the evolution operator in the combined Faraday-Ampere equations is not unitary. There are two steps needed to convert this equation into a unitary evolution equation. In the first step, a weighted Hilbert space is formulated in which the generator of dynamics is a pseudo-Hermitian operator. In the second step, a Dyson map is constructed which maps the weighted--physical--Hilbert space to the original Hilbert space. The resulting evolution equation for the electromagnetic wave fields is unitary. Utilizing the framework developed in these steps, a unitary evolution equation is derived for electromagnetic wave propagation in a uniaxial dielectric medium. The resulting form is suitable for quantum computing.
\end{abstract}


\maketitle

\section{Introduction}
The prospect that, for a range of problems, quantum computers could be be exponentially faster than conventional computers \cite{Arute,Wu} has led to an enhanced interest in quantum computer sciences. For efficient use of quantum computers, it is necessary that the evolution equations for any physical system be expressed in terms of unitary operators \cite{Nielsen}. There is no such requirement for classical computations. The tantalizing possibility of faster computations as well as including many more degrees of freedom has been the motivation behind applying quantum information science to traditionally classical fields. 

The propagation and scattering of electromagnetic waves in magneto-dielectric matter has been of considerable interest over many decades. The electromagnetic properties of a medium are included in Maxwell equations through constitutive relations that relate the electric displacement field and magnetic induction to the electric field and magnetic intensity, respectively. Most of these studies are classical -- the de Broglie wavelengths being negligibly small compared to the wavelengths of the macroscopic fields. Thus, the implementation of Maxwell equations on quantum computers requires expressing a classical description in the language of quantum mechanics. The first step in this direction was taken by Laporte-Uhlenbeck \cite{Laporte} and Oppenheimer \cite{Oppenheimer} casting Maxwell equations in vacuum into a form similar to the Dirac equation. Along similar lines, there have been recent studies drawing on the connection between the photon wave function and the Dirac equation in vacuum \cite{Smith,Mohr},  and in magneto-dielectric medium with scalar permittivity and permeability \cite{Khan}.

In this paper we formulate Maxwell equations for wave propagation in a dielectric medium such that they become amenable to quantum computations. The magneto-dielectric medium is assumed to be passive and non-dispersive for which both the permittivity and permeability can be a tensor. When the medium is spatially homogeneous, the Faraday-Ampere equations take on the form of a Dirac equation for spin 1 photons. The evolution operator is unitary and the state vector is a six-vector composed of the electric field and magnetic intensity. A unitarily similar representation is obtained for a state vector comprising of Riemann-Silberstein-Weber (RSW) vectors \cite{Birula1} -- which represent the left and right hand polarizations of an electromagnetic field.

When the same prescription is extended to a spatially inhomogeneous medium, the evolution operator for the electromagnetic fields is not unitary anymore.  We develop a pathway towards a unitary evolution equation through a two-step process. The first step is to identify the generator of dynamics -- the Hamiltonian operator -- as a pseudo-Hermitian operator. In a newly-defined weighted Hilbert space the Hamiltonian is Hermitian with respect to a weighed inner product structure.
There has been a lot of interest in pseudo-Hermitian Hamiltonians in quantum mechanics, especially in the subset of  $PT$-symmetric Hamiltonians \cite{Bender1,Bender2,Mostafazadeh1}. The second step is to draw a connection between the physical Hilbert space and the initial Hilbert space that preserves the inner product structure of the two spaces. This is accomplished by constructing an appropriate isometric Dyson map. The end result is a fully unitary evolution equation with an explicit Hermitian Hamiltonian that could be implemented in a quantum computer. The fields evaluated from this evolution equation are directly related to the physical electromagnetic fields.

This paper is organized as follows. In sections \ref{sec:2.1} and \ref{sec:2.2} we formulate the Faraday-Ampere equations in terms of a six-vector and in terms of RSW vectors, respectively, for a homogeneous medium. In section \ref{sec:2.3}, we extend the description to allow for an inhomogeneous medium. From Poynting's theorem, as expected for a passive medium, we show that the total electromagnetic energy is conserved in a bounded medium subject to suitably chosen Dirichlet boundary conditions. In section \ref{sec:2.4}, it is shown 
that the evolution generator of the previous section is not Hermitian due to spatial inhomogeneity. A physical Hilbert space is  created in which the Hamiltonian is Hermitian. It is shown that, in this weighted Hilbert space, the norm of the state vector is the conserved energy that follows from Poynting's theorem. In section \ref{sec:2.5}, we formulate three different forms of the Dyson map which lead to a 
Maxwell-Dirac equation with unitary evolution operator in the initial Hilbert space. In section \ref{sec:2.6}, the entire formalism is applied to a uniaxial dielectric medium. There is a natural extension of the evolution equation to a set of spatially dependent RSW vectors which are a generalization of the RSW vectors in \ref{sec:2.2}. In section \ref{sec:3.1}, we construct a Qubit Lattice Algorithm (QLA) corresponding to our unitary formulation of Maxwell equations. The advantage of this QLA is that it can also
be implemented and tested on classical computers. 
In section \ref{sec:3.2}, to demonstrate proof of concept, we map out a quantum circuit for the QLA that is suitable for a quantum computer.
\section{Quantum Representation}
The source-free Maxwell equations for a linear medium are,
\begin{align}
   \div{\bol{D} (\bol{r}, t)} &= 0, & \div{\bol{B}(\bol{r}, t)} &= 0, \label{max1} \\
   \pdv{\bol{B}(\bol{r}, t)}{t} &= -\curl{\bol{E}(\bol{r}, t)}, & \pdv{\bol{D}(\bol{r}, t)}{t} &= \curl{\bol{H}(\bol{r}, t)}, \label{max2}
\end{align}
with the constitutive relations,
\begin{equation}
\bol{D}(\bol{r}, t) = \epsilon (\bol{r}) \ \bol{E}(\bol{r}, t), \quad \quad \bol{B}(\bol{r}, t) = \mu(\bol{r}) \ \bol{H}(\bol{r}, t), \label{max3}
\end{equation}
where $\bol{E}$ is the electric field, $\bol{B}$ is the magnetic induction, $\bol{D}$ is the displacement field, $\bol{H}$ is the magnetic intensity, $\epsilon$ is the dielectric permittivity of the medium and  $\mu$ is its magnetic permeability; $\epsilon$ and $\mu$ can be functions of space.

In section \ref{sec:2.1}, we express Maxwell equations in terms of a six-vector when $\epsilon$ and $\mu$ are independent of space and time.
We show that the Faraday-Ampere equations \eqref{max2} take on a form similar to the Dirac equation for spin 1 massless photon. In section \ref{sec:2.2}, we rewrite the Faraday-Ampere system using the RSW vectors and draw similarities with the results in section \ref{sec:2.1}.

In section \ref{sec:2.3}, we assume that the medium is inhomogeneous in space, independent of time, and non-dissipative -- i.e., $\epsilon(\bol{r})$ and $\mu(\bol{r})$ are real functions. The Faraday-Ampere equations and the Poynting theorem are set up using the six-vector representation.

\subsection{Six-vector formulation of Maxwell equations} \label{sec:2.1}

The Faraday-Ampere equations \eqref{max2} can be written in a compact form using a six-vector \cite{Lindell},
\begin{equation}\label{9}
i\pdv{\bol{u}}{t}= \widehat{W}^{-1}\widehat{M}\bol{u}=\widehat{D}\bol{u},
\end{equation}
where $\bol{u}=(\bol{E} \ \ \bol{H})^T$ is an ordered pair of three-vectors composed of the electromagnetic field, $T$ indicates the transpose,
\begin{equation}\label{10}
\widehat{M}=i\begin{bmatrix}
0&\curl\\
-\curl&0
\end{bmatrix},\quad \quad \quad \widehat{W}=\begin{bmatrix}
\epsilon{I}_{3\times3}&0_{3\times3}\\
0_{3\times3}&\mu{I}_{3\times3}
\end{bmatrix},
\end{equation}
$I_{3\times3}$ is the $3 \times 3$ identity matrix, and $0_{3\times3}$ is the null matrix. The invertible, Hermitian matrix $\widehat{W}$ operating on $\bol{u}$ yields the constitutive relations $\bol{D}(\bol{r}, t) = \epsilon \bol{E}(\bol{r}, t)$ and $\bol{B}(\bol{r}, t) = \mu \bol{H}(\bol{r}, t)$ for a homogeneous medium.
The Maxwell operator $\widehat{M}$ is Hermitian in $L^2(\mathbb{R}^3,\mathbb{C})$ with the appropriate boundary conditions. We will discuss this further in sections \ref{sec:2.3} and \ref{sec:2.4}. 

The  generator of the evolution operator $\widehat{D}$ in \eqref{9} is Hermitian since the Hermitian operators $\widehat{W}^{-1}$ and $\widehat{M}$ commute, $\widehat{W}^{-1}\widehat{M}=\widehat{M}\widehat{W}^{-1}$. Upon operating on \eqref{9} with $\widehat{W}^{1/2}$, we obtain,
\begin{equation}\label{11}
 i\pdv{\bol{U}}{t}=(-\sigma_y\otimes{v}\bol{S}\cdot\widehat{\bol{p}})\bol{U}=\widehat{D}_{\rho}\bol{U},
\end{equation}
where $v=1/\sqrt{\epsilon\mu}$ is the speed of light in the medium, $\bol{U}=\widehat{W}^{1/2}\bol{u}$, the components of $\bol{S} = \left( S_x, S_y, S_z \right)$ are the spin 1 matrices,
\begin{equation}\label{5}
{S}_x=\begin{bmatrix}
0&0&0\\
0&0&-i\\
0&i&0
\end{bmatrix}\quad{S_y}=\begin{bmatrix}
0&0&i\\
0&0&0\\
-i&0&0
\end{bmatrix}\quad{S_z}=\begin{bmatrix}
0&-i&0\\
i&0&0\\
0&0&0
\end{bmatrix},
\end{equation}
satisfying the commutator relation $[S_a,S_b]=i\epsilon_{abc}S_c$, $\widehat{\bol{p}}=-i \nabla$ is equivalent to the quantum momentum operator for $\hbar = 1$, and the Pauli spin 1/2 matrices $\bol{\sigma} = \left( \sigma_x, \sigma_y, \sigma_z \right)$ are,
\begin{equation}\label{pauli}
{\sigma}_x=\begin{bmatrix}
0&1\\
1&0
\end{bmatrix}\quad{\sigma_y}=\begin{bmatrix}
0&-i\\
i&0
\end{bmatrix}\quad{\sigma_z}=\begin{bmatrix}
1&0\\
0&-1
\end{bmatrix}.
\end{equation}
Since the evolution operator $\widehat{D}_\rho$ is Hermitian, \eqref{11} is analogous to the Dirac equation for a spin 1 massless photon.

 \subsection{Riemann-Silberstein-Weber vectors and Maxwell equations }\label{sec:2.2}
 
The RSW vectors $\bol{F}^{\pm} (\bol{r},t)$ are a re-expression of the electromagnetic fields in a form that is useful for a quantum-like formulation of Maxwell equations. They are defined as \cite{Birula1},
 \begin{equation}\label{3}
\bol{F}^\pm(\bol{r},t) =\frac{1}{\sqrt{2}}\Big(\sqrt{\epsilon} \, \bol{E} \pm \frac{i}{\sqrt{\mu}} \, \bol{B} \Big).
\end{equation}
For a homogeneous medium, 
the Faraday-Ampere equations take on the form \cite{Good,Birula1},
\begin{equation}\label{4}
i\pdv{\bol{F}^\pm}{t} = \pm v(\bol{S}\cdot\widehat{\bol{p}})\bol{F}^\pm.
\end{equation}
Equation \eqref{4} can be considered as a quantum representation of Maxwell equations with the RSW vectors as the photon wave function \cite{Birula1}.

In the standard square-integrable Hilbert space $\mathcal{H}=L^2(\mathbb{R}^3,\mathbb{C})$, the Hermitian Hamiltonian operator in \eqref{4},
\begin{equation}\label{6}
\widehat{H}=v(\bol{S}\cdot\widehat{\bol{p}}),
\end{equation}
has eigenvalues $E=\omega$, reflecting the monochromatic energy of a photon. This is analogous to the quantum definition of energy for $\hbar = 1$.  The norm of the RSW vectors $\bol{F}^\pm $ 
is the electromagnetic energy of the macroscopic electromagnetic field,
\begin{widetext}
\begin{equation}\label{7}
\braket{\bol{F}^\pm}=\norm{\bol{F}^\pm}^2=\int_\Omega\bol{F}^{\pm \dagger}\bol{F^\pm }\ d\,\bol{r}=\frac{1}{2}\int_\Omega\Big(\epsilon\bol{E}^2+\frac{\bol{B}^2}{\mu}\Big)d\,\bol{r},
\end{equation}
\end{widetext}
where $\dagger$ is the complex conjugate transpose of the vector.

The evolution equation \eqref{4} is analogous to the Weyl equation for spin 1/2 massless particles,
\begin{equation}\label{8}
i\pdv{\bol{\psi}}{t}=c(\bol{\sigma}\cdot\widehat{\bol{p}})\bol{\psi},
\end{equation}
where $\bol{\psi}$ is the wave function composed of the two Weyl spinors. The analogy is not surprising since the two RSW vectors represent the two distinct polarizations of the electromagnetic field in a homogeneous, time-independent medium.

If we introduce a unitary transformation $\widehat{L}:\bol{U}\to\bol{F}$ where,
\begin{equation}\label{12}
\widehat{L}=\frac{1}{\sqrt{2}}\begin{bmatrix}
I_{3\times3}&i\, I_{3\times3}\\
I_{3\times3}&-i\, I_{3\times3}
\end{bmatrix},
\end{equation}
then \eqref{11} takes the block diagonal form,
\begin{equation}\label{13}
i\pdv{}{t}\begin{bmatrix}
\bol{F^+}\\
\bol{F^-}
\end{bmatrix}=\begin{bmatrix}
v\bol{S}\cdot\widehat{\bol{p}}&0\\
0&-v\bol{S}\cdot\widehat{\bol{p}}
\end{bmatrix}\begin{bmatrix}
\bol{F^+}\\
\bol{F^-}
\end{bmatrix},
\end{equation}
which is exactly the form in \eqref{4}. The six-vector form of the Faraday-Ampere equations is directly connected to the RSW vectors. In other words, the RSW transformation is a Weyl representation of the Dirac-type equation \eqref{11}. Significantly, the representations \eqref{11} and \eqref{13} are equivalent due to the unitary nature of the transformation \eqref{12}. In a homogeneous medium, the two field helicities are uncoupled as is the time evolution of the RSW vectors.

\subsection{Maxwell equations in an inhomogeneous, passive medium} \label{sec:2.3}

The Faraday-Ampere equations \eqref{max2} can be written as,
\begin{equation}\label{extra3}
i\pdv{\bol{d}}{t}=\widehat{M}\bol{u}.
\end{equation}
where $\bol{d}(\bol{r},t)=(\bol{D},\bol{B})^T$ is related to $\bol{u}(\bol{r},t)=(\bol{E},\bol{H})^T$ by a linear constitutive operator $\widehat{\mathcal{L}}$,
\begin{equation}\label{14}
\bol{d}=\bol{d}(\bol{u})\Rightarrow\bol{d}=\widehat{\mathcal{L}}\bol{u}.
\end{equation}
The divergence equations \eqref{max1} become,
\begin{equation}
\div \bol{d} = \div{ \left( \widehat{\mathcal{L}} \bol{u} \right)} = 0. \label{diveq}
\end{equation}
If at time $t = 0$, $\bol{d}_0 = \bol{d} \left( \bol{r}, 0 \right)$ is such that,
\begin{equation}\label{15}
\div\bol{d}_0=\div \left( \widehat{\mathcal{L}}\bol{u}_0 \right) = 0, 
\end{equation}
where $\bol{u}_0 = \bol{u} \left( \bol{r}, 0 \right)$, then
\eqref{extra3} ensures that $\div{ \bol{d} \left( \bol{r}, t \right)} = 0 $
is for all times. We will assume that the medium is bounded by a perfect conductor so that,
\begin{equation}\label{17}
\widehat{\bol{n}}(\bol{r})\times\bol{u}_1=0\,\,{\rm on\,\, the \,boundary}\,\,\partial\Omega,
\end{equation}
where $\widehat{\bol{n}}(\bol{r})$ is the outward pointing normal at the boundary, and $\bol{u}= \left( \bol{u}_1, \bol{u}_2 \right)^T$ with $\bol{u}_1$ and $\bol{u}_2$ each being a three-vector. The boundary conditions are necessary for energy conservation and for ensuring that $\widehat{M}$ remains Hermitian.
The set of equations \eqref{extra3}-\eqref{17} are the complete mathematical description of electromagnetic waves in a Hilbert state space $\mathcal{H}=L^2(\Omega,\mathbb{R}^6)\ni\bol{u}$ defined by the inner product \cite{Roach},
\begin{equation}\label{18}
\braket{\bol{v}}{\bol{u}}=\int_{\Omega}\bol{v}^\dagger(\bol{r},t)\bol{u}(\bol{r},t)d\,\bol{r},\quad\Omega\subseteq\mathbb{R}^3,\,\,t\in\mathcal{T}=[0,T],
\end{equation}
where $\bol{u} \left( \bol{r}, t \right)$ and $\bol{v} \left( \bol{r}, t \right)$ are two solutions within the bounded domain defined by $\Omega$.

A general form of the constitutive operator  $\widehat{\mathcal{L}}$ has to satisfy five physical postulates \cite{Roach}: determinism, linearity, causality, locality in space, and invariance under time translations. The form that is consistent with these postulates is \cite{Roach},
\begin{equation}\label{19}
\bol{d}(\bol{r},t) = \widehat{\mathcal{L}}\bol{u}(\bol{r},t)= \widehat{W}(\bol{r})\bol{u}(\bol{r},t)+\int_0^t\widehat{G}(\bol{r},t-\tau)\bol{u}(\bol{r},\tau)d\,\tau.
\end{equation}
The first term on the right hand side in \eqref{19} corresponds to instantaneous optical response of the medium, and the second term with $G$ as the susceptibility kernel is the dispersive response which includes memory effects. 

For an anisotropic non-dispersive medium, \eqref{19} reduces to,
\begin{equation}
\bol{d}(\bol{r},t) = \widehat{W}(\bol{r})\bol{u}(\bol{r},t), \label{nondisp}
\end{equation}
where,
\begin{equation}\label{20}
\widehat{W}=\begin{bmatrix}
\epsilon(\bol{r})&0_{3\times3}\\
0_{3\times3}&\mu(\bol{r})
\end{bmatrix}.
\end{equation}
In what follows, we will ignore dispersive effects. However, in general, a constitutive relation of the form \eqref{nondisp} is an approximation to \eqref{19} which includes a non-local time-response function \cite{Landau}. 

Since $\widehat{W}$ is invertible \cite{Roach}, \eqref{extra3} takes on the form,
\begin{equation}\label{22}
i\pdv{\bol{u}}{t}= \widehat{W}^{-1}(\bol{r})\widehat{M}\bol{u}=\widehat{D}\bol{u},
\end{equation}
where,
\begin{equation}
\widehat{D} = \widehat{W}^{-1}(\bol{r}) \widehat{M}. \label{ddef}
\end{equation}
In this representation, the Poynting theorem is \cite{Roach},
\begin{equation}\label{23}
\div{\bol{S}}+\bol{u}^\dagger\pdv{\bol{d}}{t}=0,
\end{equation}
where $\bol{S}=\bol{E}\times\bol{H}$ is the Poynting vector.
Following \cite{Friden}, the electromagnetic energy density is,
\begin{widetext}
\begin{equation}\label{24}
U (\bol{r},t)=\int_0^t\bol{u}^\dagger \ \pdv{\bol{d}(\bol{r},\tau)}{\tau} \ d\tau = \frac{1}{2}\bol{u}^\dagger{\widehat{W}}\bol{u}+\int_0^t\bol{u}^{\dagger}(\bol{r},\tau)\ \widehat{W}^A \ \pdv{\bol{u}(\bol{r},\tau)}{\tau} \ d\tau,
\end{equation}
\end{widetext}
where $\widehat{W}^A= \left( \widehat{W}- \widehat{W} ^\dagger \right) / 2$ is the anti-Hermitian part of $\widehat{W}$.
For a passive medium \cite{Roach,Friden},
\begin{equation}\label{25}
U (\bol{r},t)\geq0,\quad\forall \ \bol{r}\in\Omega.
\end{equation}
From \eqref{24} it follows that $\widehat{W}$  must be Hermitian and semi-positive definite,
\begin{equation}\label{26}
\widehat{W} = \widehat{W}^\dagger \quad\quad {\rm and} \quad \quad \widehat{W} \geq 0.
\end{equation}
The total integrated stored electromagnetic energy $U_\Omega$ in a volume $\Omega$  is, 
\begin{equation}\label{27}
U_{\Omega} (t)=\frac{1}{2}\ \int_\Omega\bol{u}^\dagger \ \widehat{W} \ \bol{u} \ d^3\bol{r} \ \geq \ 0.
\end{equation}
Integrating \eqref{23} over $\Omega$ and making use of the divergence theorem, we obtain,
\begin{equation}\label{28}
\pdv{U_\Omega(t)}{t}+\int_{\partial\Omega}\ \bol{S}\cdot\widehat{\bol{n}}(\bol{r}) \ dA = 0,
\end{equation}
where $dA$ is an elemental area on the surface $\partial \Omega$, and $\widehat{\bol{n}}$ is the outward pointing normal to $\partial \Omega$.
Since $\bol{S} = \bol{u}_1 \times \bol{u}_2$, it follows from \eqref{28} and the boundary condition \eqref{17} that $U_\Omega (t)$ is constant in time, 
\begin{equation}\label{33}
{U_\Omega}(t) = U_\Omega (t=0) = \int_\Omega\bol{u}_0^\dagger \,\widehat{W}(\bol{r}) \, \bol{u}_0 \ d^3\bol{r}.
\end{equation}
As expected for a passive medium, there is no net dissipation or generation of electromagnetic energy within $\Omega$.

\subsection{Pseudo-Hermitian Operators \label{sec:2.4}}

In the Hilbert space $\mathcal{H}$,
\begin{widetext}
\begin{subequations}\label{30}
\begin{align}
\braket{\bol{v}}{\widehat{M}\bol{u}}&=i\int_{\Omega}\Big(\bol{v}^*_1\cdot(\curl{\bol{u}_2})-\bol{v}^*_2\cdot(\curl{\bol{u}_1})\Big) \ d^3\bol{r} \label{30a}\\ 
&=i\int_{\Omega} \Big( \bol{u}_2\cdot(\curl{\bol{v}^*_1})-\bol{u}_1\cdot(\curl{\bol{v}^*_2}) \Big) \ d^3\bol{r}+
i\int_{\partial\Omega}\Big(\bol{v}^*_1\cdot( \widehat{\bol{n}} \times \bol{u}_2 )- \bol{v}^*_2\cdot \left( \widehat{\bol{n}}\times\bol{u}_1 \right)\Big) \ dA, \label{30b}
\end{align}
\end{subequations}
\end{widetext}
where $\bol{v} = \left( \bol{v}_1, \bol{v_2} \right)^T$, with $\bol{v}_1$, and $\bol{v}_2$ each being three-vectors.
In obtaining \eqref{30b} from \eqref{30a}, we have made use of the vector identity $\div{(\bol{a}\times\bol{b})}=\bol{b}\cdot(\curl{\bol{a}})-\bol{a}\cdot(\curl{\bol{b}})$, and the divergence theorem. Since $\bol{v}_1^* \cdot \left( \widehat{\bol{n}} \times \bol{u}_2 \right) = - \bol{u}_2 \cdot \left( \widehat{\bol{n}} \times \bol{v}_1^* \right) $, the surface integral in \eqref{30b} vanishes as a consequence of the boundary condition \eqref{17}. It follows from \eqref{30} that,
\begin{equation}\label{31}
\braket{\bol{v}}{\widehat{M}\bol{u}} = \braket{\bol{v}\widehat{M}}{\bol{u}},
\end{equation}
proving that $\widehat{M}$ is Hermitian; i.e., $\widehat{M}=\widehat{M}^\dagger$.

Even though $\widehat{W}(\bol{r})$ and $\widehat{M}$ are Hermitian, the operator $\widehat{D}=\widehat{W}^{-1}\widehat{M}$ is not Hermitian. In contrast to a homogeneous medium, the commutator $\left[ \widehat{W}^{-1},\widehat{M} \right]$ is non-zero for an inhomogeneous medium. Consequently, the operator on the right-hand side of the evolution equation \eqref{22} is non-unitary. In order to make Maxwell equations  for an inhomogeneous medium suitable for quantum computing, we formulate a unitary representation that relies on $\widehat{D}$ being a special kind of non-Hermitian operator -- a pseudo-Hermitian operator. 

A linear operator $\widehat{D}$ in a Hilbert space $\mathcal{H}$ is pseudo-Hermitian if there exists an invertible Hermitian linear operator $\widehat{\eta}$ in $\mathcal{H}$ with the property \cite{Mostafazadeh2,Znojil2,Fring},
 \begin{equation}\label{34}
\widehat{D}^{\dagger}=\widehat{\eta} \, \widehat{D} \, {\widehat{\eta}}^{-1}.
 \end{equation}
 For $\mel{\bol{u}}{\widehat{\eta}}{\bol{u}}> 0$ for all nonzero states $\bol{u}$, $\widehat{\eta}$ is a positive definite metric operator, and we can define an inner product,
 \begin{equation}\label{35}
\braket{\bol{v}}{\bol{u}}_\eta=\braket{\bol{v}}{\widehat{\eta}\bol{u}}
= \int_\Omega\bol{v}^\dagger(\bol{r},t) \, \widehat{\eta} (\bol{r}) \, \bol{u}(\bol{r},t)\ d^3\bol{r},
\end{equation}
with respect to a new weighted Hilbert space $\mathcal{H}_\eta$. 

From \eqref{ddef}, 
\begin{equation}\label{36}
\widehat{D}^\dagger = \widehat{M}\, \widehat{W}^{-1} = \widehat{W}\, \widehat{W}^{-1} \, \widehat{M} \, \widehat{W}^{-1} = \widehat{W}\, \widehat{D} \, \widehat{W}^{-1}.
\end{equation}
Comparing with \eqref{34} we note that $\widehat{\eta}=\widehat{W}$. Furthermore, exploiting the Hermicity condition \eqref{35} of Maxwell operator $\widehat{M}$ we obtain,
\begin{equation}
\begin{aligned}
\braket{\bol{v}}{\widehat{D}\bol{u}}_W&=\braket{\bol{v}}{\widehat{M}\bol{u}}=\braket{\bol{v}\widehat{M}\widehat{W}^{-1}\widehat{W}}{\bol{u}}\\
&=\braket{\bol{v}\widehat{M}\widehat{W}^{-1}}{\bol{u}}_W=\braket{\bol{v}\widehat{D}}{\bol{u}}_W. \label{eta2}
\end{aligned}
\end{equation}
Thus, $\widehat{D}^\dagger = \widehat{D}$, i.e., $\widehat{D}$ is Hermitian in the weighted Hilbert space $\mathcal{H}_W$. In $\mathcal{H}_W$, the inner product is as defined in \eqref{35} with $\widehat{\eta}$ replaced by $\widehat{W}$, and the evolution equation \eqref{22} is unitary. Making use of \eqref{33}, the square of the norm of $\bol{u}$,
\begin{equation}\label{39}
\braket{\bol{u}}_W = \mel{\bol{u}}{\widehat{W}}{\bol{u}} = \int_\Omega\ \bol{u}^\dagger(\bol{r},t)\, 
\widehat{W}(\bol{r}) \, \bol{u}(\bol{r},t)\ d^3\bol{r} = 2 \, U_{\Omega},
\end{equation}
is a constant independent of time. Consequently, the underlying conservation of the electromagnetic energy in the closed volume $\Omega$ is preserved in $\mathcal{H}_W$.

\subsection{The Dyson map for  Maxwell equations \label{sec:2.5}}

Even though $\widehat{D}$ is Hermitian in the new Hilbert space $\mathcal{H}_{W}$, in the Maxwell-Dirac equation \eqref{22} there is no change except that $\bol{u}\in\mathcal{H}_{W}$. We need to connect the original, physical, Hilbert space $\mathcal{H}$, in which $\widehat{D}$ is not Hermitian, to $\mathcal{H}_{W}$ by an isometric transformation that preserves the inner product structure between the two Hilbert spaces. Such an invertible transformation $\widehat{\rho}(\bol{r}):\mathcal{H}_{W}\to\mathcal{H}$ between equivalent descriptions of a physical system is referred to as a Dyson map \cite{Znojil2,Fring,Mostafazadeh2}.

Our derivation of the Dyson map $\widehat{\rho}(\bol{r})$ is based on the factorization of the metric operator $\widehat{\eta}=\widehat{W}$,
\begin{equation}\label{41}
\widehat{\eta}(\bol{r})=\widehat{\rho}^\dagger(\bol{r})\widehat{\rho}(\bol{r}),
\end{equation}
which preserves the inner product structure since,
\begin{equation}\label{42}
\braket{\bol{v}}{\bol{u}}_\eta=\braket{\bol{v}}{\widehat{\rho}^\dagger\widehat{\rho}\bol{u}}=\braket{\bol{v}\widehat{\rho}}{\widehat{\rho}\bol{u}}=\braket{\bol{\phi}}{\bol{\psi}},
\end{equation}
where $\bol{v},\bol{u}\in\mathcal{H}_{W}$ and $\bol{\phi},\bol{\psi}\in\mathcal{H}$. 

For Maxwell equations, there can be three different factorization forms of $\widehat{W}$ in \eqref{20} \cite{Horn}.
\begin{itemize}
    \item {\bf Spectral decomposition}:
\begin{equation}\label{43}
\widehat{W}(\bol{r})=\widehat{U}^\dagger{\widehat{\Delta}}(\bol{r})\widehat{U}=\widehat{U}^\dagger\sqrt{\widehat{\Delta}(\bol{r})}\sqrt{\widehat{\Delta}(\bol{r)}}\widehat{U}=\widehat{\rho}^\dagger(\bol{r})\widehat{\rho}(\bol{r}),
\end{equation}
leading to the Dyson map,
\begin{equation} \label{dyson1}
\widehat{\rho}(\bol{r})=\sqrt{\widehat{\Delta}(\bol{r})}\widehat{U}.
\end{equation}

\item {\bf Square root decomposition}:
\begin{equation}\label{44}
\widehat{W}(\bol{r})= \widehat{W}^{1/2}(\bol{r}) \widehat{W}^{1/2}(\bol{r})=\widehat{\rho}^\dagger(\bol{r})\widehat{\rho}(\bol{r}),
\end{equation}
with the corresponding Dyson map,
\begin{equation} \label{dyson2}
\widehat{\rho}(\bol{r})= \widehat{W}^{1/2}(\bol{r}).
\end{equation}

\item {\bf Cholesky decomposition}:
\begin{equation}\label{45}
\widehat{W}(\bol{r})=\widehat{T}^\dagger(\bol{r})\widehat{T}(\bol{r})=\widehat{\rho}^\dagger(\bol{r})\widehat{\rho}(\bol{r}),
\end{equation}
giving the Dyson map,
\begin{equation}
\quad\widehat{\rho}(\bol{r})=\widehat{T}(\bol{r}).
\end{equation}

\end{itemize}
 For the spectral decomposition \eqref{43}, $\widehat{\Delta}(\bol{r})=\lambda_i(\bol{r})\delta_{ij}$ (there is no implied summation over repeated indices) with $\lambda(\bol{r})>0$ and $\sqrt{\widehat{\Delta}(\bol{r})}=\sqrt{\lambda_i(\bol{r})}\delta_{ij}$. For the Cholesky decomposition \eqref{45}, the $\widehat{T}$ matrix is an upper triangular matrix with positive diagonal elements. The particular choice of a Dyson map is based on the decomposition scheme which leads to a sparse $\widehat{\rho}$.

The Dyson map leads to a Hermitian form for Maxwell equations in $\mathcal{H}$. Multiplying \eqref{22} by $\widehat{\rho}$ gives,
\begin{equation}\label{46}
i\pdv{\bol{\psi}}{t}=\widehat{\rho}(\bol{r})\widehat{D}\widehat{\rho}^{-1}(\bol{r})\bol{\psi}=\widehat{D}_{\rho}\bol{\psi},
\end{equation}
where $\bol{\psi}=\widehat{\rho}\bol{u}\in\mathcal{H}$, and $\widehat{D}_\rho = \widehat{\rho}(\bol{r})\widehat{D}\widehat{\rho}^{-1}(\bol{r})$ is Hermitian in $\mathcal{H}$
The unitary evolution of $\bol{\psi} \left( \bol{r}, t \right)$ is,
\begin{equation}\label{47}
\bol{\psi}(\bol{r},t)=e^{-it\widehat{D}_\rho} \bol{\psi}_0 \left( \bol{r} \right),
\end{equation}
where $\bol{\psi}_0 \left( \bol{r} \right)$ is the initial condition at time $t=0$.

\begin{figure}[h]
\centering
\[\begin{tikzcd}
    \bol{u}\in\mathcal{H}_\eta\ar{r}{\widehat{\rho}}\ar{d}[swap]{\text{definition of}\,\, \mathcal{H}_\eta} & 
    \ar{ld}{\widehat{\rho}^{-1}}\bol{\psi}\in\mathcal{H} \\
    \bol{u}\in\mathcal{H}.
\end{tikzcd}\]
\caption{The Dyson map interconnection between the various spaces.} \label{fig:1}
\end{figure}
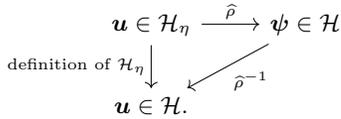

In general, any operator $\widehat{A}_\eta:\mathcal{H}_\eta\to\mathcal{H}_{\eta}$ is related to its counterpart $\widehat{A}$ in $\mathcal{H}$ through a similarity transformation,
\begin{equation}\label{48}
\widehat{A} = \widehat{\rho} \, \widehat{A}_\eta\, \widehat{\rho}^{-1}.
\end{equation}
The Dyson map $\widehat{\rho}$ connecting  $\mathcal{H}$ to $\mathcal{H}_\eta$ can be schematically represented in Figure ~\ref{fig:1}.
This diagram illustrates the dual role of the Dyson map. The first is to map the weighted space $\mathcal{H}_\eta$ into the initial Hilbert space $\mathcal{H}$ through an isometric transformation. This ensures that the Hamiltonian is Hermitian and the evolution is unitary.  The second is to map different, not equivalent, representation of elements $\bol{u}$ and $\bol{\psi}$ belonging to $\mathcal{H}$. This is evident from the Dyson mapping -- the operator $\widehat{\rho}:\mathcal{H}\to\mathcal{H}$ is not unitary in $\mathcal{H}$. In other words, the transformation $\widehat{\rho}:\bol{u}\to\bol{\psi}$ is not a trivial and unitary representation of the initial dynamics \eqref{22}. However, every other transformation $\widehat{\tau}$ in $\mathcal{H}$ that preserves the dynamics of  \eqref{46} is unitary. From  \eqref{46}, applying the transformation $\widehat{\tau}$, the generator $\widehat{D}_\tau$ yields,
\begin{equation}\label{unitary}
\widehat{D}_\tau=\widehat{\tau}\widehat{D}_\rho\widehat{\tau}^{-1}=\widehat{D}^\dagger_{\tau}=(\widehat{\tau}^{-1})^\dagger\widehat{D}_\rho\widehat{\tau}^\dagger\Rightarrow\widehat{\tau}^{-1}=\widehat{\tau}^\dagger.
\end{equation}
Thus, all other dynamics preserving transformations are equivalent once the Dyson map $\widehat{\rho}$ is established. Indeed, this holds for the formulation in terms of RSW  vectors $\bol{F}=\widehat{L}\widehat{W}^{1/2}\bol{u}$.

\subsection{Application to a uniaxial dielectric medium \label{sec:2.6}}

As an illustration of the formalism developed in section \ref{sec:2.5}, we consider a non-magnetic, uniaxial dielectric medium,
\begin{equation}\label{50}
\epsilon(\bol{r})=\begin{bmatrix}
\epsilon_x(\bol{r})&0&0\\
0&\epsilon_x(\bol{r})&0\\
0&0&\epsilon_z(\bol{r})
\end{bmatrix},\quad \quad \quad \mu=\mu_0I_{3\times3}.
\end{equation}
A useful choice for a sparse Dyson map is,
\begin{equation}\label{51}
\widehat{\rho}=\widehat{W}^{1/2}=\begin{bmatrix}
\epsilon^{1/2}&0_{3\times3}\\
0_{3\times3}&\sqrt{\mu_0}I_{3\times3},
\end{bmatrix},
\end{equation}
where,
\begin{equation} \label{51a}
\epsilon^{1/2} \left( \bol{r} \right) = \begin{bmatrix}
\sqrt{\epsilon_x(\bol{r})}&0&0\\
0&\sqrt{\epsilon_x(\bol{r})}&0\\
0&0&\sqrt{\epsilon_z(\bol{r})},
\end{bmatrix}.
\end{equation}
Then, $\widehat{D}_\rho = \widehat{\rho} \, \widehat{D} \, \widehat{\rho}^{-1} $ is,
\begin{equation}\label{52}
\widehat{D}_\rho = \begin{bmatrix}
0_{3\times3}&ic\bol{Z}\cdot\widehat{\bol{p}}\\
-ic\widehat{\bol{p}}\cdot\bol{Z}^\dagger&0_{3\times3}
\end{bmatrix},
\end{equation}
where, in terms of the refractive index $n_i(\bol{r})=\sqrt{\epsilon_i(\bol{r}) / {\epsilon_0}}$, the components of $\bol{Z}=(Z_x,Z_y,Z_z)$ are
\begin{widetext}
\begin{equation}\label{53}
Z_x=\begin{bmatrix}
0&0&0\\
0&0&- \displaystyle{\frac{i}{n_x \left( \bol{r} \right) }} \\
0& \displaystyle{\frac{i}{n_z}\left( \bol{r} \right)}&0
\end{bmatrix},\quad{Z}_y=\begin{bmatrix}
0&0&\displaystyle{\frac{i}{n_x\left( \bol{r} \right)}}\\
0&0&0\\
- \displaystyle{\frac{i}{n_z\left( \bol{r} \right)}}&0&0
\end{bmatrix},\quad{Z}_z=\begin{bmatrix}
0&- \displaystyle{\frac{i}{n_x\left( \bol{r} \right)}}&0\\
\displaystyle{\frac{i}{n_x\left( \bol{r} \right)}}&0&0\\
0&0&0
\end{bmatrix},
\end{equation}
\end{widetext}
with $n_x = \sqrt{ \epsilon_x / \epsilon_0 }$ and $n_z = \sqrt{ \epsilon_z / \epsilon_0 }$ being the indices of refraction in the $x$ and $z$ directions, respectively, and, as before, $\widehat{\bol{p}} = - i\nabla$.

Applying the unitary operator $\widehat{L}$ in \eqref{12} to \eqref{46} gives,
\begin{equation}
i \pdv{}{t} \widehat{L}\bol{\psi} = \left( \widehat{L} \widehat{D}_\rho\widehat{L}^{-1} \right) \, \widehat{L}\bol{\psi}. \label{newf1}
\end{equation}
Upon defining,
\begin{equation}\label{55}
\bol{F}_{r}^{\pm} \ = \ \widehat{L}\bol{\psi} \ = \ \widehat{L} \widehat{W}^{1/2}\bol{u} \ = \ \frac{1}{\sqrt{2}} \, \Big(\epsilon^{1/2} (\bol{r}) \, \bol{E} \, \pm \, \frac{i}{\sqrt{\mu_0}} \, \bol{B} \Big),
\end{equation}
the unitary evolution equation \eqref{newf1} takes on the form,
\begin{equation}\label{56}
i\pdv{}{t}\begin{bmatrix}
\bol{F}_{r}^+\\
\bol{F}_{r}^-
\end{bmatrix}=c\begin{bmatrix}
(\bol{Z}\cdot\widehat{\bol{p}})^H&-(\bol{Z}\cdot\widehat{\bol{p}})^A\\
(\bol{Z}\cdot\widehat{\bol{p}})^A&-(\bol{Z}\cdot\widehat{\bol{p}})^H
\end{bmatrix}
\begin{bmatrix}
\bol{F}_{r}^+\\
\bol{F}_{r}^-
\end{bmatrix},
\end{equation}
where the superscripts $H$ and $A$ represent the Hermitian and the anti-Hermitian parts of the operator, respectively. The definition in \eqref{55} for an inhomogeneous medium is a generalization of the RSW vectors \eqref{3} for a homogeneous medium. For a non-dissipative medium, the anti-Hermitian part of $\bol{Z}$ in \eqref{56} is zero, and the time evolution of the two RSW vectors $\bol{F}^\pm$ decouples. It is straightforward to show that for a homogeneous non-dissipative medium, \eqref{56} reduces to \eqref{13}.
% If in two-column mode, this environment will change to single-column
% format so that long equations can be displayed. Use
%\begin{widetext}
%\end{widetext}

\section{Connection with Quantum Computing}\label{sec:3}

The representation of Maxwell equations expressed in
  \eqref{46} is suitable for implementing on a quantum computer as it
  satisfies a primary requirement -- unitarity. In
  addition, the operator $\widehat{D}_\rho$ for Maxwell equations is
  equivalent to any other Hermitian representation within
  $\mathcal{H}$, as Eq. \eqref{unitary} suggests. Consequently, any
  algorithm developed for implementation on quantum computers for the
  unitary operator $\exp{it\widehat{D}_\rho}$ also applies to any
  other unitary evolution of the same system. This particular aspect regarding the equivalence of two unitary operators within the same physical Hilbert space  is also discussed in \cite{Croke} where the Dyson map is referred to as a ``passive transformation''. However, it is important to note that we need to have an explicit form for $\widehat{D}_\rho$ in order to take advantage of quantum computing. In the next subsection, we develop a qubit lattice algorithm (QLA) for $\widehat{D}_\rho$ in a bi-axial dielectric medium which is suitable for implementing on a quantum computer.

\subsection{Qubit Lattice Algorithms}\label{sec:3.1}

Qubit lattice algorithms have been used to simulate the propagation and scattering of electromagnetic waves is an inhomogeneous dielectric medium having a scalar permittivity \cite{Vahala1,Ram,Vahala2}. A QLA is a discrete representation of Maxwell equations, usually up to second order in a perturbation parameter, which, at a mesoscopic level, uses an appropriately chosen interleaved sequence of three non-commuting operators. Two of the operators are collision and streaming operators -- the collision operator entangles the on-site qubits and the streaming operator propagates the entangled state through the lattice. The dielectric medium is included via a third operator referred to as a potential operator. 
Following \cite{unpublished}, we construct a QLA for two-dimensional scattering of electromagnetic waves by a bi-axial dielectric material described by a diagonal refractive index $n(\bol{r})=diag(n_x,n_y,n_z)$.

Following the discussion in section \ref{sec:2.6}, the state vector that admits unitary evolution has the form,
\begin{equation}\label{eq:61}
\begin{bmatrix}
n_xE_x\\
n_yE_y\\
n_zE_z\\
\mu_0^{1/2}H_x \\
\mu_0^{1/2}H_y \\
\mu_0^{1/2}H_z
\end{bmatrix}=\begin{bmatrix}
q_0\\
q_1\\
q_2\\
q_3\\
q_4\\
q_5
\end{bmatrix}=\bol{q}.
\end{equation}
Assuming two-dimensional spatial dependence in the $x$-$y$ plane, the decomposition of the optical Dirac equation \eqref{46} into Cartesian components yields,
\begin{equation}\label{62}
\begin{aligned}
&\pdv{q_0}{t}=\frac{1}{n_x}\pdv{q_5}{y},\quad \pdv{q_1}{t}=\frac{1}{n_y}\pdv{q_5}{y},\quad \pdv{q_2}{t}=\frac{1}{n_z}\Big[\pdv{q_4}{y}-\pdv{q_3}{x}\Big],\\
&\pdv{q_3}{t}=\pdv{(q_2/n_z)}{y},\quad \pdv{q_4}{t}=\pdv{(q_2/n_z)}{x},\\
&\pdv{q_5}{t}=-\pdv{(q_1/n_y)}{x}+\pdv{(q_0/n_x)}{n_y}.
\end{aligned}
\end{equation}
We discretize the two-dimensional space into a lattice with the spacing given by the ordering parameter $\mathcal{O}(\delta)$. Then, to second order in $\delta$,
the unitary collision operators in the $x$ and $y$ directions are, respectively,
\begin{equation}\label{63}
\widehat{C}_X=\begin{bmatrix}
 1&0&0&0&0&0\\
 0&\cos{\theta_1}&0&0&0&-\sin{\theta_1}\\
 0&0&\cos{\theta_2}&0&-\sin{\theta_2}&0\\
 0&0&0&1&0&0\\
 0&0&\sin{\theta_2}&0&\cos{\theta_2}&0\\
 0&\sin{\theta_1}&0&0&0&\cos{\theta_1}
\end{bmatrix},
\end{equation}
\begin{equation}\label{64}
\widehat{C}_Y=\begin{bmatrix}
 \cos{\theta_0}&0&0&0&0&\sin{\theta_0}\\
 0&1&0&0&0&0\\
 0&0&\cos{\theta_2}&\sin{\theta_2}&0&0\\
 0&0&-\sin{\theta_2}&\cos{\theta_2}&0&0\\
 0&0&0&0&1&0\\
 -\sin{\theta_0}&0&0&0&0&\cos{\theta_0}
\end{bmatrix}.
\end{equation}
Let $\widehat{S}_{ij}$ denote a unitary streaming operator which shifts the qubits $q_i$ and $q_j$ one lattice unit along $x$ and one lattice along $y$,  
while leaving all the other qubits unaffected. Then the collide-stream sequence along each direction is,
\begin{widetext}
\begin{equation}\label{65}
\begin{aligned}
\widehat{U}_X&=\widehat{S}^{+x}_{25}\widehat{C}^\dagger_X\widehat{S}^{-x}_{25}\widehat{C}_X\widehat{S}^{-x}_{14}\widehat{C}^\dagger_X\widehat{S}^{+x}_{14}\widehat{C}_X\widehat{S}^{-x}_{25}\widehat{C}_X\widehat{S}^{+x}_{25}\widehat{C}^\dagger_X\widehat{S}^{+x}_{14}\widehat{C}_X\widehat{S}^{-x}_{14}\widehat{C}^\dagger_X\\
\widehat{U}_Y&=\widehat{S}^{+y}_{25}\widehat{C}^\dagger_Y\widehat{S}^{-y}_{25}\widehat{C}_Y\widehat{S}^{-y}_{03}\widehat{C}^\dagger_Y\widehat{S}^{+y}_{03}\widehat{C}_Y\widehat{S}^{-y}_{25}\widehat{C}_Y\widehat{S}^{+y}_{25}\widehat{C}^\dagger_Y\widehat{S}^{+y}_{03}\widehat{C}_Y\widehat{S}^{-y}_{03}\widehat{C}^\dagger_Y.
\end{aligned}
\end{equation}
\end{widetext}
The terms in \eqref{62} that contain the derivatives of the refractive index are recovered through the following potential operators,
\begin{equation}\label{66}
\widehat{V}_X=\begin{bmatrix}
1&0&0&0&0&0\\
0&1&0&0&0&0\\
0&0&1&0&0&0\\
0&0&-\sin{\beta_2}&0&\cos{\beta_2}&0\\
0&\sin{\beta_0}&0&0&0&\cos{\beta_0}
\end{bmatrix}
\end{equation}
and
\begin{equation}\label{67}
\widehat{V}_Y=\begin{bmatrix}
1&0&0&0&0&0\\
0&1&0&0&0&0\\
0&0&1&0&0&0\\
0&0&\cos{\beta_3}&\sin{\beta_3}&0&0\\
-\sin{\beta_1}&0&0&0&0&\cos{\beta_1}
\end{bmatrix}.
\end{equation}
The  angles $\theta_0$, $\theta_1$, $\theta_2$, $\beta_0$, $\beta_1$, $\beta_2$, and $\beta_3$ that appearing in \eqref{63}, \eqref{64}, \eqref{66}, and \eqref{67} are chosen so that the discretized system reproduces \eqref{62} to order $\delta^2$.
The evolution of the state vector $\bol{q}$ from time $t$ to $t+\Delta{t}$ is given by,
\begin{equation}\label{68}
\bol{q}(t+\Delta{t})=\widehat{V}_Y\widehat{V}_X\widehat{U}_Y\widehat{U}_X\bol{q}(t).
\end{equation}
The external potential operators $\widehat{V}_X, \widehat{V}_Y$, as given above, are not unitary. Nonetheless, we can implement 
$\widehat{V}_{X,Y}$ in our algorithm using the method of linear combinations of unitary operators (LCU) \cite{Childsnew,Childsnew2}.

\subsection{Quantum Encoding}\label{sec:3.2}
The product decomposition formula describing the evolution of the state $\bol{q}$ in \eqref{68}, form the core of quantum simulation \cite{Nielsen}. 
An efficient quantum algorithm requires all the unitary evolution operators to be encoded into simple quantum gates. For this, we construct two qubit registers --
the first for encoding the amplitude of the state vector $\bol{q}$, and the second for the discrete $x$-$y$ space. SInce the state vector $\bol{q}$ is six-dimensional, 
the first register will contain $n_i=3$ qubits with basis $\ket{i}$ and amplitudes $q_i$. For the two-dimensional lattice with $N$ nodes and a discretization step $\delta$ in both 
directions, we will need $n_p=\log_2N$ qubits with basis $\ket{p}$. Hence, we will need $n_{total}=n_p+3$ qubits for a complete description of state $\bol{q}$. 
The qubit encoding of the state vector $\bol{q}$ on a lattice site is,
\begin{equation}\label{69}
\ket{\bol{q}}=\sum_{i=0}^{5}q_i\ket{i}\ket{p},
\end{equation}
where the amplitudes $q_i$ are normalized to the square root of the initial (constant) energy $U_{\Omega}(0)$ in \eqref{33},  so that $\sum_i\abs{q_i}^2=1$.

\subsubsection{Preparation of initial state}\label{subsec:1}
Preparation of initial state $\ket{\bol{q}_0}$, made up of real $6N$-components, is expressed in terms of 
the amplitudes of a quantum state using a sequence of controlled one-qubit rotations,
\begin{equation}\label{stateprep1}
\ket{000}\ket{0}^{\otimes{n_p}}\to\ket{q_0}.
\end{equation}
In general, this requires a quantum circuit of $\mathcal{O}(6N)$ elementary gates. 
However, for studying propagation and scattering of electromagnetic waves in physically relevant situations, the initial state, like wave-packets or pulses, is localized in space. 
Thus, the initial condition will usually be a small subset of the complete $N$-dimensional discretized space,
\begin{equation}\label{stateprerp2}
\ket{\bol{q}_0}=\sum_{p=0}^{M}\sum_{i}q_{0ip}\ket{i}\ket{p}
\end{equation}
where $q_{0ip}=0$ for $p>M$ and $M<<N$.
The sparse initial state \eqref{stateprerp2} uses $\mathcal{O}(6M)$ gates, thereby reducing the overall cost of implementation.

\subsubsection{Implementation of \texorpdfstring{$\widehat{C}_{X,Y}$}{TEXT} operators}\label{subsec:2}
\begin{figure}[h]
\includegraphics{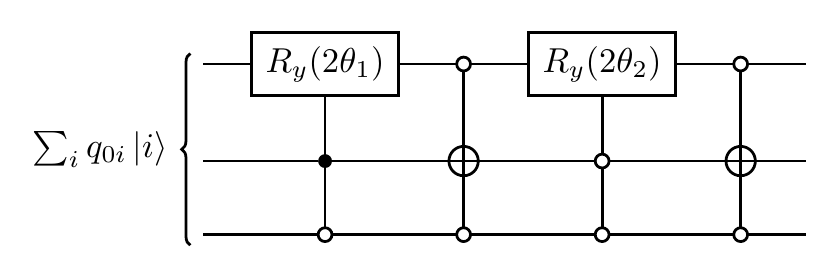}
%\begin{quantikz}
%\lstick[wires=3]{$\sum_{i}{q_{0i}}\ket{i}$} 
%& \gate{R_y(2\theta_1)} & \octrl{2} & \gate{R_y(2\theta_2)} & \octrl{1} & \qw \\
 %& \ctrl{-1} & \targ{} & \octrl{-1} & \targ{} & \qw \\
 %& \octrl{-2} & \octrl{-2} & \octrl{-1} & \octrl{-1} &\qw
%\end{quantikz}
    \caption{Quantum gate implementation of $\widehat{C}_X$ acting on the $\ket{i}$ register. The $R_y$ gate corresponds to a rotation around the $y$-axis.}
    \label{fig:2}
\end{figure}

\begin{figure}[h]
\includegraphics[scale=0.9]{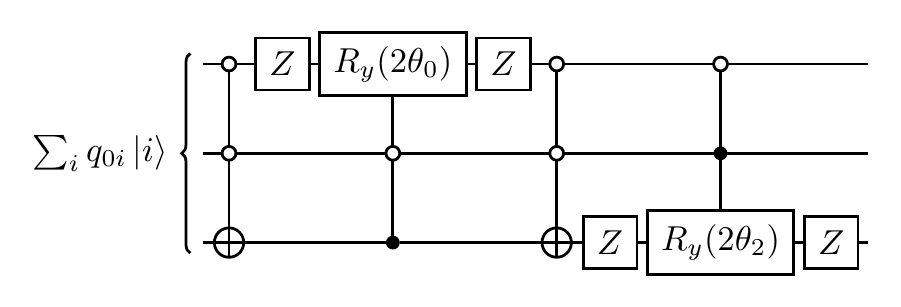}
%\begin{quantikz}[ column sep=0.3em]
%\lstick[wires=3]{$\sum_{i}{q_{0i}}\ket{i}$}
%& \octrl{1} & \gate{Z} & \gate{R_y(2\theta_0)} & \gate{Z} & \octrl{1} & \qw & \octrl{2} & \qw &\qw  \\ 
%& \octrl{1} & \qw & \octrl{-1} & \qw & \octrl{1} & \qw & \ctrl{-1} & \qw & \qw\\
 %& \targ{} & \qw &\ctrl{-1} &\qw & \targ{} & \gate{Z} & \gate{R_y(2\theta_2)} & \gate{Z} & \qw
%\end{quantikz}
    \caption{Quantum gate implementation of $\widehat{C}_Y$ acting on the $\ket{i}$ register. The $Z$ gate corresponds to the Pauli matrix $\sigma_z$.}
    \label{fig:3}
\end{figure}

We assign the unitary collision operators $\widehat{C}_X$ in \eqref{63} and $\widehat{C}_Y$ in \eqref{64}  to multi-controlled, single-qubit unitary gates. 
Since these operators act on $\ket{i}$, we obtain the following two-level unitary decomposition,
\begin{widetext}
\begin{equation}\label{70}
\begin{aligned}
\widehat{C}_X&=\begin{bmatrix}
 1&0&0&0&0&0\\
 0&\cos{\theta_1}&0&0&0&-\sin{\theta_1}\\
 0&0&1&0&0&0\\
 0&0&0&1&0&0\\
 0&0&0&0&1&0\\
 0&\sin{\theta_1}&0&0&0&\cos{\theta_1}
\end{bmatrix}
\begin{bmatrix}
 1&0&0&0&0&0\\
 0&1&0&0&0&0\\
 0&0&\cos{\theta_2}&0&-\sin{\theta_2}&0\\
 0&0&0&1&0&0\\
 0&0&\sin{\theta_2}&0&\cos{\theta_2}&0\\
 0&0&0&0&0&1
\end{bmatrix}\\
\widehat{C}_Y&=\begin{bmatrix}
 \cos{\theta_0}&0&0&0&0&\sin{\theta_0}\\
 0&1&0&0&0&0\\
 0&0&1&0&0&0\\
 0&0&0&1&0&0\\
 0&0&0&0&1&0\\
 -\sin{\theta_0}&0&0&0&0&\cos{\theta_0}
\end{bmatrix}\begin{bmatrix}
1&0&0&0&0&0\\
 0&1&0&0&0&0\\
 0&0&\cos{\theta_2}&\sin{\theta_2}&0&0\\
 0&0&-\sin{\theta_2}&\cos{\theta_2}&0&0\\
 0&0&0&0&1&0\\
0&0&0&0&0&1
\end{bmatrix}.
\end{aligned}
\end{equation}
\end{widetext}
Subsequently, the quantum gate implementation of $\widehat{C}_X$ and $\widehat{C}_Y$ acting on $\ket{i}$ is as depicted in Figures \ref{fig:2} and \ref{fig:3}, respectively.

\begin{figure}[h]
\includegraphics{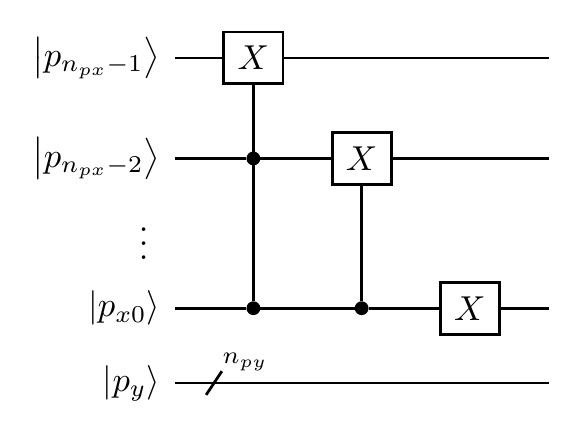}
%\begin{quantikz}
%& \lstick{$\ket{p_{n_{px}-1}}$} & \gate{X} & \qw & \qw & \qw \\
%& \lstick{$\ket{p_{n_{px}-2}}$} & \ctrl{-1} & \gate{X} & \qw & \qw \\
%& \lstick{\myvdots} &&&\\
%& \lstick{$\ket{p_{x0}}$}  & \ctrl{-2} & \ctrl{-2} &\gate{X} &\qw \\
%& \lstick{$\ket{p_y}$}  &\qwbundle
%{n_{py}} &\qw & \qw & \qw
%\end{quantikz}
\caption{Quantum gate implementation of streaming operator $\widehat{S}^{+x}$ in the $\ket{p}$ register. The least significant bit is the $p_{x0}$.}
\label{fig:4}
\end{figure}

\begin{figure}[h]
\includegraphics{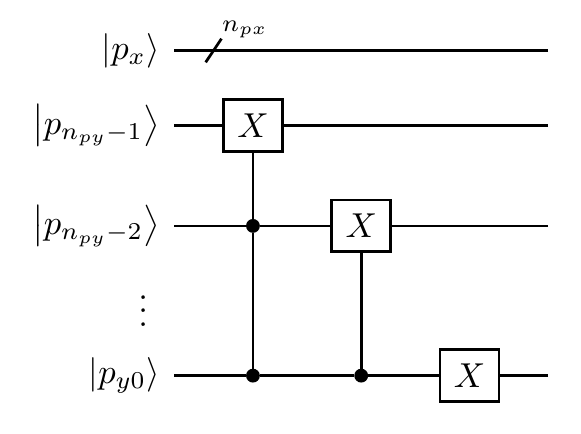}
%\begin{quantikz}
%& \lstick{$\ket{p_x}$}  &\qwbundle
%{n_{px}} &\qw & \qw & \qw \\
%& \lstick{$\ket{p_{n_{py}-1}}$} & \gate{X} & \qw & \qw & \qw \\
%& \lstick{$\ket{p_{n_{py}-2}}$} & \ctrl{-1} & \gate{X} & \qw & \qw \\
%& \lstick{\myvdots} &&&\\
%& \lstick{$\ket{p_{y0}}$}  & \ctrl{-2} & \ctrl{-2} &\gate{X} &\qw 
%\end{quantikz}
\caption{Quantum gate implementation of streaming operator $\widehat{S}^{+y}$ in $\ket{p}$. The least significant bit is $p_{y0}$.}
    \label{fig:5}
\end{figure}

For the two-dimensional lattice with $N = N_x N_y$ nodes, there are $N_x - 1$ and $N_y - 1$ number of segments of length $\delta$
along each direction. Consequently, the $\ket{p}$ register contains two sub-registers for each spatial direction
with $n_{px}$ and ${n_{py}}$ number of qubits along $x$ and $y$, respectively. Thus,
\begin{equation}\label{71}
n_{p}=\log_2{N}=\log_2{N_x}+\log_2{N_y}=n_{px}+{n_{py}}.
\end{equation}
The spatial location of each node is given by,
\begin{equation}\label{72}
\ket{p}=\ket{p_x}\ket{p_y}=\ket{a_x+p_x\delta}\ket{a_y+p_y\delta},
\end{equation}
with $p_x=0, 1, ..., N_x-1$ and $p_y=0, 1, ..., N_y-1$. The action of streaming operators on the $\ket{p}$ register,
\begin{equation}\label{73}
\widehat{S}^{+x}\ket{p}=\ket{p_x+1}\ket{p_y},\quad \widehat{S}^{+y}\ket{p}=\ket{p_x}\ket{p_y+1},
\end{equation}
is controlled by the qubits in the $\ket{i}$ register as is evident from the sequence in \eqref{65}.
Expressing $\ket{p}$ in its binary form $\ket{p_{n_{px}-1}p_{n_{px}-2}...p_{x0}}\ket{p_{n_{py}-1}p_{n_{py}-2}...p_{y0}}$, 
the implementation of \eqref{73}  is shown in Figures \ref{fig:4} and \ref{fig:5}. For simplicity, the control dependence on $\ket{i}$ has been omitted.
Following \eqref{73}, the action of $\widehat{S}^{-x},\widehat{S}^{-y}$ is represented using the conjugate transpose quantum circuit, 
since $\widehat{S}^{-x,-y}=(\widehat{S}^{+x,+y})^\dagger$.

\subsubsection{LCU operations for  \texorpdfstring{$\widehat{V}_{X,Y}$}{TEXT}operators}\label{subsec:3}
The sparse operators $\widehat{V}_{X,Y}$ in \eqref{66}, \eqref{67} can be decomposed into a 4-term unitary sum,
\begin{equation}\label{76}
\widehat{V}_{X,Y}=\frac{1}{2}\sum_{j=0}^4\widehat{\Tilde{V}}_{j_{X,Y}},
\end{equation}
where the unitary matrices $\widehat{\Tilde{V}}_j$ are,
\begin{widetext}
\begin{equation}\label{77}
\begin{aligned}
\widehat{\Tilde{V}}_{0X}&=\widehat{\Tilde{V}}_{0Y}=I_{6\times6}\\
\widehat{\Tilde{V}}_{1X}&=\begin{bmatrix}
-1&0&0&0&0&0\\
0&1&0&0&0&0\\
0&0&1&0&0&0\\
0&0&0&-1&0&0\\
0&0&0&0&-1&0\\
0&0&0&0&0&-1
\end{bmatrix},\quad \widehat{\Tilde{V}}_{1Y}=\begin{bmatrix}
1&0&0&0&0&0\\
0&-1&0&0&0&0\\
0&0&1&0&0&0\\
0&0&0&-1&0&0\\
0&0&0&0&-1&0\\
0&0&0&0&0&-1
\end{bmatrix}\\
\widehat{\Tilde{V}}_{2X}&=\begin{bmatrix}
 1&0&0&0&0&0\\
 0&\cos{\beta_0}&0&0&0&-\sin{\beta_0}\\
 0&0&\cos{\beta_2}&0&\sin{\beta_2}&0\\
 0&0&0&1&0&0\\
 0&0&-\sin{\beta_2}&0&\cos{\beta_2}&0\\
 0&\sin{\beta_0}&0&0&0&\cos{\beta_0}
\end{bmatrix},\quad \widehat{\Tilde{V}}_{2Y}=\begin{bmatrix}
 \cos{\beta_1}&0&0&0&0&\sin{\beta_1}\\
 0&1&0&0&0&0\\
 0&0&-\sin{\beta_3}&\cos{\beta_3}&0&0\\
 0&0&\cos{\beta_3}&\sin{\beta_3}&0&0\\
 0&0&0&0&1&0\\
 -\sin{\beta_1}&0&0&0&0&\cos{\beta_1}
\end{bmatrix}\\
\widehat{\Tilde{V}}_{3X}&=\begin{bmatrix}
 1&0&0&0&0&0\\
 0&-\cos{\beta_0}&0&0&0&\sin{\beta_0}\\
 0&0&-\cos{\beta_2}&0&-\sin{\beta_2}&0\\
 0&0&0&1&0&0\\
 0&0&-\sin{\beta_2}&0&\cos{\beta_2}&0\\
 0&\sin{\beta_0}&0&0&0&\cos{\beta_0}
\end{bmatrix},\quad \widehat{\Tilde{V}}_{3Y}=\begin{bmatrix}
 -\cos{\beta_1}&0&0&0&0&-\sin{\beta_1}\\
 0&1&0&0&0&0\\
 0&0&\sin{\beta_3}&-\cos{\beta_3}&0&0\\
 0&0&\cos{\beta_3}&\sin{\beta_3}&0&0\\
 0&0&0&0&1&0\\
 -\sin{\beta_1}&0&0&0&0&\cos{\beta_1}
\end{bmatrix}.
\end{aligned}
\end{equation}
\end{widetext}
As a result, the evolution operator $\widehat{U}_{ev}$ in Eq. \eqref{68} is a sum of unitary operators,
\begin{equation}\label{78}
\widehat{U}_{ev}=\frac{1}{4}\sum_{j,k}^3\widehat{\Tilde{V}}_{jX}\widehat{\Tilde{V}}_{kY}\widehat{U}_X\widehat{U}_Y=\Big(\sum_{m=0}^{15}\widehat{U}_m\Big)\widehat{U}_X\widehat{U}_Y.
\end{equation}

In order to implement \eqref{78}, we need to apply the LCU method. For an ancillary register  of $n_m=\log_216=4$ qubits we define the following unitary operators,
\begin{equation}\label{79}
\begin{aligned}
\widehat{U}_{select}&=\sum_{m=0}^{15}\ket{m}\bra{m}\otimes\widehat{U}_m\\
 \widehat{U}_{prep}&: \ket{0}^{\otimes{n_m}}\to\frac{1}{4}\sum_{m=0}^{15}\ket{m},
\end{aligned}
\end{equation}
where $\widehat{U}_{prep}$ is the state preparation operator in the ancillary register. The implementation of $\widehat{U}_{select}$ is similar to that in Figs. \ref{fig:2}-\ref{fig:3} ; $\widehat{U}_m$ are composed of dual combinations of two-level matrices \eqref{77}
containing rotations and Pauli gates.

Finally,  following \cite{Childsnew2}), we implement $\widehat{U}_{ev}$ 
using $\widehat{W}=\widehat{U}_{prep}\widehat{U}_{select}\widehat{U}^\dagger_{prep}$ where,
\begin{equation}\label{80}
\widehat{W}(\ket{0}^{\otimes{n_m}}\widehat{U}_X\widehat{U}_Y\ket{\bol{q}_0})=\frac{1}{4}\ket{0}^{\otimes{n_m}}\ket{\bol{q}}+\ket{\Psi^{\perp}},
\end{equation}
with $(\ket{0}^{\otimes{n_m}}\prescript{{n_m}\otimes}{}{\bra{0}}\otimes1)\ket{\Psi^{\perp}}=0$.
A measurement in the ancillary register leads to the desired outcome with probability $1/16$.

\subsubsection{Discussion}\label{subsec:4}
The quantum circuits in Figs. \ref{fig:2}-\ref{fig:5} along a representation of the initial state, fully implement the unitary sequences 
$\widehat{U}_X$ and $\widehat{U}_Y$ in \eqref{65} using $\mathcal{O}[16M(n_{total}+2)]$ multi-controlled single qubit gates. For $M=\mathcal{O}(n^{\kappa}_p)<<N$, we can reduce the number of gates to $ \mathcal{O}(n^{\kappa+1}_p)$. By introducing an ancillary register of $n_m=4$ qubits,
the implementation cost of $\widehat{U}_{prep}$ and $\widehat{U}_{select}$, using LCU, scales as $\mathcal{O}(12M+16)=\mathcal{O}(n_p^{\kappa})$. 
Consequently, the number of multi-controlled single qubit gates that are needed to effectively simulate \eqref{68} is $\Theta(n_p^{\kappa+1})$. 
The polynomial gate complexity of a simulation depends primarily on the qubit number $n_p$ associated with spatial discretization.

Finally, retrieval of physically relevant information (electromagnetic energy, $\bol{E}$ and $\bol{B}$ fields) from the final state can be achieved by employing proper projection operators and amplitude estimation \cite{Brassard}.

\section{Conclusions}
The propagation and scattering of electromagnetic waves in a dielectric medium is governed by
classical Maxwell equations. The enticing possibility of an exponential reduction in computational
time on quantum computers has led to recasting some topics in classical physics into the framework
of quantum information science. We have expressed the Faraday-Ampere equations for a passive, non-
dispersive dielectric medium in a form that is similar to the Dirac equation for a massless spin 1 particle.
For a medium homogeneous in space, the permittivity and permeability are scalars. In the Dirac-
type evolution equation, the electromagnetic fields are expressed either as a six-vector or as Riemann-
Silberstein-Weber vectors. When the permittivity and permeability of the medium are functions of
space then, in contrast to a homogeneous medium, the Maxwell operator $\widehat{M}$ and the operator for the
constitutive relations $\widehat{W}$ do not commute. Even though both operators are Hermitian, their product
is not. A remedy is to construct a weighted Hilbert space $\mathcal{H}_W$ in which the generator of dynamics 
is pseudo-Hermitian. The Hermitian, positive definite metric operator that defines the inner product within 
$\mathcal{H}_W$ is $\widehat{W}$. The norm of a state vector in $\mathcal{H}_W$  is the electromagnetic 
energy which, from Poynting’s theorem, is conserved. The connection between the original Hilbert space 
and $\mathcal{H}_W$ is established through a Dyson map.  Significantly, the Dyson map is a fundamental 
way to construct a unitary evolution of Maxwell equations for wave propagation in a complex medium. Any 
other representation, preserving the dynamics, is generated through a unitary transformation of the Dyson map. 
There are three different Dyson maps that are suitable for connecting the two Hilbert spaces. The preferred Dyson map could be 
guided by the sparseness of the associated matrix operators. Regardless of the choice, the final form of the Faraday-Ampere 
equations comprises unitary evolution operators.
The formal development of Maxwell equations into a unitary evolution equation is applied to a
uniaxial, inhomogeneous, dielectric medium. 
We use a qubit lattice algorithm to illustrate a means of implementing our formalism on to a 
quantum computer. The backbone of the QLA, which uses the state variable $\bol{q}$ as its qubit basis, 
is an interleaved sequence of unitary collision and streaming operators. The
collision operators entangle the on-site qubits, while the streaming operators move this entanglement throughout the lattice. In contrast to the 
Lie-Trotter-Suzuki treatment of non-commuting exponential Hermitian operators, the QLA consists of sparse matrices. The present formulation of 
QLA needs external potential operators which are sparse but not unitary. However, following \cite{Childsnew,Childsnew2}, the potential
operators can be represented as a sum of unitary operators making them amenable for quantum computers.
Consequently, as we have shown, it is possible to design the appropriate quantum circuits.
Since the polynomial gate complexity of a simulation scales with 
the number  $n_p$ of qubits which, in turn, are related to the number of spatial grid points, the speedup of quantum computing 
for simulation with high spatial resolution is quite clear. Even though it is early to estimate the scale of the speedup for
QLA algorithms, recent developments provide reasons for optimism. It is likely that optimized QLA will make use of fast Fourier transform techniques
for implementing the streaming operators, thereby reducing the number of operations of the streaming operator
\cite{Oganesov}.

\begin{acknowledgments}
This work has been carried out within the framework of the EUROfusion Consortium, funded by the European Union via the Euratom Research and Training Programme (Grant Agreement No 101052200 — EUROfusion). Views and opinions expressed are however those of the authors only and do not necessarily reflect those of the European Union or the European Commission. Neither the European Union nor the European Commission can be held responsible for them.
A.K.R is supported by the Department of Energy under Grant Nos. DE-SC0021647 and DE-FG02-91ER-54109.
G.V is supported by the Department of Energy under Grant Nos. DE-SC0021651.
\end{acknowledgments}

\providecommand{\noopsort}[1]{}\providecommand{\singleletter}[1]{#1}%

\end{document}